\documentclass[conference]{IEEEtran}

\ifCLASSINFOpdf
\usepackage[pdftex]{graphicx}
\else
\fi

\usepackage[cmex10]{amsmath}

\usepackage{url}

\newenvironment{definition}[1][Definition]{\begin{trivlist}
\item[\hskip \labelsep {\bfseries #1}]}{\end{trivlist}}

\begin{document}
\title{Clustering Memes in Social Media}

\author{\IEEEauthorblockN{Emilio Ferrara$^{1,*}$, Mohsen JafariAsbagh$^1$, Onur Varol$^1$, Vahed Qazvinian$^2$, Filippo Menczer$^1$, Alessandro Flammini$^1$}
\IEEEauthorblockA{\small $^1$Center for Complex Networks and Systems Research, School of Informatics and Computing, Indiana University, Bloomington, USA\\
$^2$Department of Electrical Engineering and Computer Science, University of Michigan, USA\\
$^*$Corresponding author. Address: 919 E. 10th St., Room 322A, Bloomington IN 47408 (USA), +1(812)856-7841. E-mail: ferrarae@indiana.edu}}

\maketitle

\begin{abstract}
The increasing pervasiveness of social media creates new opportunities to study human social behavior, while challenging our capability to analyze their massive data streams. One of the emerging tasks is to distinguish between different kinds of activities, for example engineered misinformation campaigns versus spontaneous communication. Such detection problems require a formal definition of \emph{meme}, or unit of information that can spread from person to person through the social network. Once a meme is identified, supervised learning methods can be applied to classify different types of communication. The appropriate granularity of a meme, however, is hardly captured from existing entities such as tags and keywords. Here we present a framework for the novel task of detecting memes by clustering messages from large streams of social data. We evaluate various similarity measures that leverage content, metadata, network features, and their combinations. We also explore the idea of pre-clustering on the basis of existing entities. A systematic evaluation is carried out using a manually curated dataset as ground truth. Our analysis shows that pre-clustering and a combination of heterogeneous features yield the best trade-off between number of clusters and their quality, demonstrating that a simple combination based on pairwise maximization of similarity is as effective as a non-trivial optimization of parameters. Our approach is fully automatic, unsupervised, and scalable for real-time detection of memes in streaming data.
\end{abstract}


%\IEEEpeerreviewmaketitle

%#####################
\section{Introduction} 
\label{sec:introduction}

The amount of information shared on online social media has been growing at unprecedented rates during recent years. Platforms such as Twitter are used for spreading news and opinions \cite{kwak2010twitter,bakshy2011everyone}, coordinating social protest efforts \cite{conover2013geospatial,conover2013digital}, %\cite{skoric2011online,morales2012users}, 
aggregating individuals with common interests \cite{wu2011says}, and more. The uncontrolled nature of social media makes them vulnerable to exploitation for spreading spam, rumors, slander, and other types of misinformation~\cite{chew2010pandemics,metaxas2010obscurity}. In the domain of politics, the more subtle phenomenon of \emph{astroturfing} has received attention in the recent literature \cite{ratkiewicz2011detecting,ratkiewicz2011truthy}. Astroturfing arises when one or a few individuals make a coordinated effort to create the false impression of a spontaneous (\emph{grassroot}) movement, inducing users to deem the information reliable and feed its propagation. 
%An effective way to achieve this goal is to produce variants of the same message, imitating the behavior of a crowd that paraphrases the same concept in different forms.  This viral spreading of astroturfing makes it nearly impossible to distinguish from organic topics once it has diffused to a large audience. It is thus of  paramount importance to develop the ability to identify such abuses early. 

The detection of these kinds of orchestrated campaigns is becoming a key problem, and a challenging one. It requires the ability to classify the massive amount of content continuously produced on online social media. Manual labeling is infeasible on a large scale. The task is also difficult due to the limitations of the textual content typical of online social media. For example, Twitter enforces a maximum length of publishable messages (\emph{tweets}) of 140 characters. Therefore, we postulate that the unit of classification should not be a single tweet, but rather a \emph{meme}, defined as a unit of information --- an idea or a concept --- that can spread from person to person through the social network. Equivalently, we can think of a meme as the \emph{set of tweets} carrying the same piece of information. Once a meme is identified, supervised learning methods can be applied to classify different types of communication. 

\begin{figure*} \centering
	\includegraphics[width=.8\textwidth]{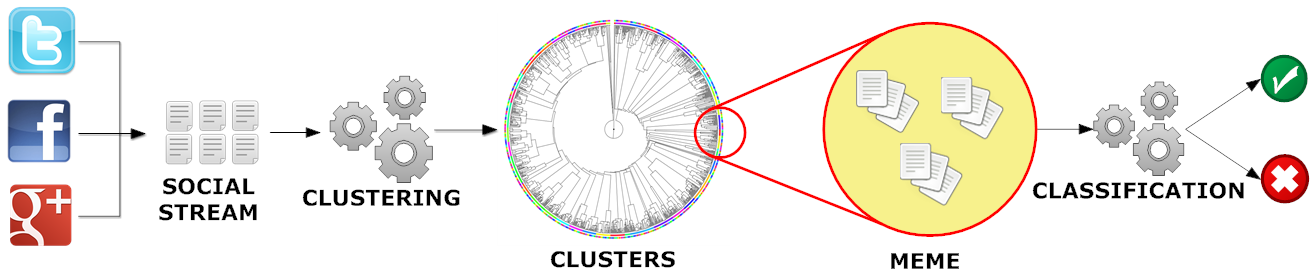}
	\caption{DESPIC architecture for meme clustering and classification}
	\label{fig:architecture}
\end{figure*}

We are developing a platform for Detecting Early Signatures of Persuasion in Information Cascades (\emph{DESPIC}), whose architecture is depicted in Fig.~\ref{fig:architecture}. There are two core components: a message clustering algorithm that takes a stream of tweets and groups them into memes, and a meme classification algorithm that labels these memes according to categories of interest. In this paper we focus on the clustering framework using Twitter as a test-bed scenario for our analysis. 

Classic document clustering techniques based on lexical analysis alone are again ineffective due to the sparsity of text, the limited context of individual tweets, and the use of references to external content. We therefore propose a strategy that leverages various sources of available metadata in addition to text. Tweets may contain \emph{hashtags}, informally defined textual tokens that are used to identify topics of discussion; \emph{mentions} of other users that are used to address messages to their attention and help identify the contributors of a conversation; and \emph{URLs} that point to external resources. We can easily group messages based on these atomic entities, that we call \emph{protomemes}. 

None of these entities alone, however, is necessarily capable to capture a meme at the appropriate level of granularity; a protomeme can be too specific or too general as a concept. Furthermore, each protomeme may only capture a particular aspect of a conversation, while a meme may require a more nuanced description. We argue that  combinations of protomemes may provide meaningful signatures of memes. Operationally, we propose a pre-clustering step based on protomemes. 

\subsection*{Contributions and outline}

This paper formalizes the problem of meme clustering and proposes an operational definition of \emph{memes} as overlapping clusters of related tweets, aggregated according to content- and network-based features. In the remainder of the paper we make the following contributions.

\begin{itemize}

\item We introduce the notion of \emph{protomemes} that provide an effective way to pre-cluster messages in real-time, streaming social media scenarios.% (\S~\ref{sub:protomemes}).

\item We define several similarity measures between memes, leveraging various content, metadata, and network features of tweets; and we propose different ways to combine them for clustering memes.% (\S~\ref{sub:measures}, \ref{sub:combinations}).

\item We compare multiple clustering algorithms, finding that 
%in the specific task of rumor detection, 
hierarchical clustering outperforms K-means in terms of the trade-off between quality of the clusters and their number and size.% (\S~\ref{sub:algorithms}).

\item We show that pre-clustering based on protomemes is an effective strategy compared to clustering the original tweets. % (\S~\ref{sub:exp-protomemes}) 

\item Finally, we compare different similarity measures and their combinations. We show that a simple combination based on pairwise maximization of similarity is as effective as a non-trivial optimization of parameters for robust performance. % (\S~\ref{sub:parameter_space}) 
Our algorithm outperforms baseline methods, including one that exploits full information about the underlying social network.

\end{itemize}

%####################
\section{Clustering framework} 
\label{sec:methodology}

The meme clustering problem is defined for any social media platform used to spread messages on a directed social network; microblogging systems like Twitter, Google Plus, and Yahoo! Meme are popular examples. In these systems, users are connected by directed links: using Twitter terminology, one \emph{follows} others to see their messages. Users can re-post (\emph{retweet}) any seen post (\emph{tweet}), spreading it to their \emph{followers}.  

In the following we introduce the notion of \emph{protomeme} and describe how protomemes have been incorporated into our clustering methodology.

\subsection{Defining protomemes} 
\label{sub:protomemes}

Let us define a set of features that can be easily extracted from a tweet and used to at least partially identify the topic of the tweet~\cite{ratkiewicz2011detecting,ratkiewicz2011truthy}:

\begin{description}

\item[Hashtag:] Twitter users can incorporate in the text of their tweets one or more \emph{hashtags}, textual tokens prefixed by hash marks (\#), that identify the topic of the message. 
%\footnote{This feature, initially not incorporated in Twitter, has emerged from users broad adoption, and now is one of the key components of the platform.}.

\item[Mention:] We say that a tweet \emph{mentions} a user when it includes the target's username preceded by the `@' symbol, thus addressing that specific user.
%\footnote{Also mentions have been incorporated after being spontaneously introduced by users.}

\item[URL:] Tweets may include links to external sources of information. A \emph{URL} is the Web address identifying a linked resource. 

\item[Phrase:] The textual content of a tweet that remains after removing hashtags, mentions, URLs, stop words, and punctuation, and after stemming words~\cite{Porter80}, is defined as a \emph{phrase}. Phrases may capture semantically equivalent lexical variations of textual messages.

\end{description}

\begin{figure*}[t!]
\centerline{
\includegraphics[height=2.2in]{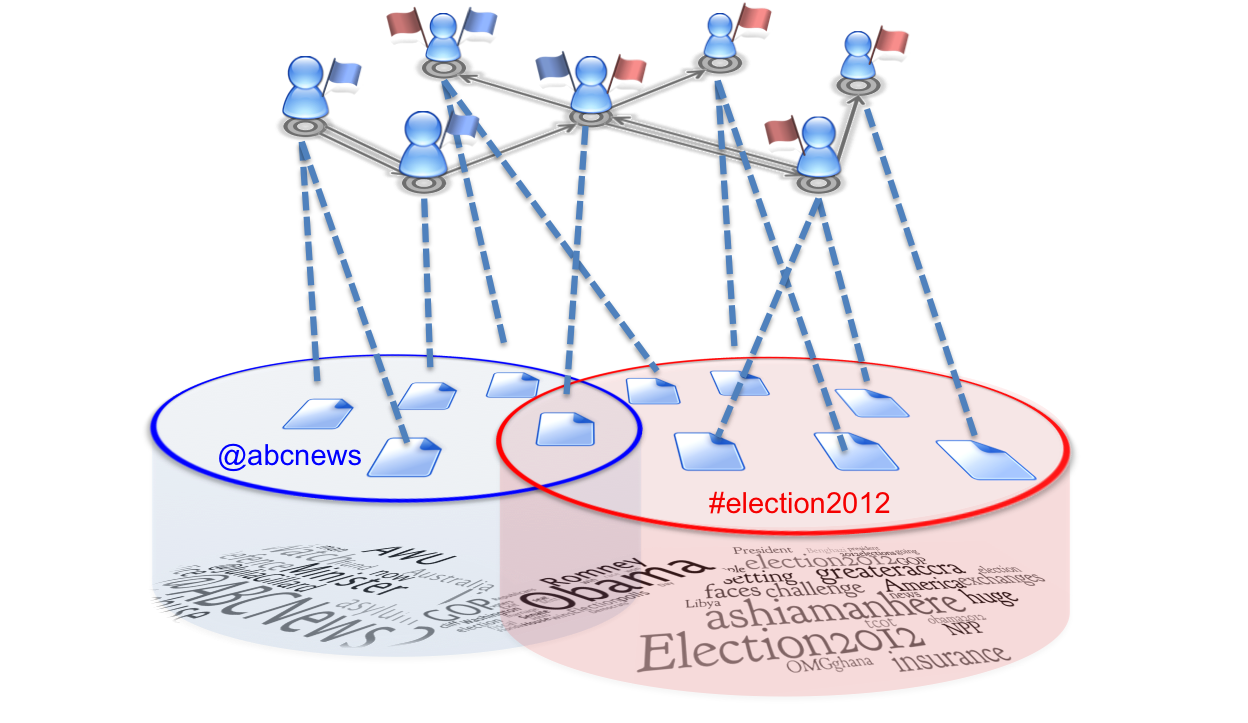}
\includegraphics[height=2.2in]{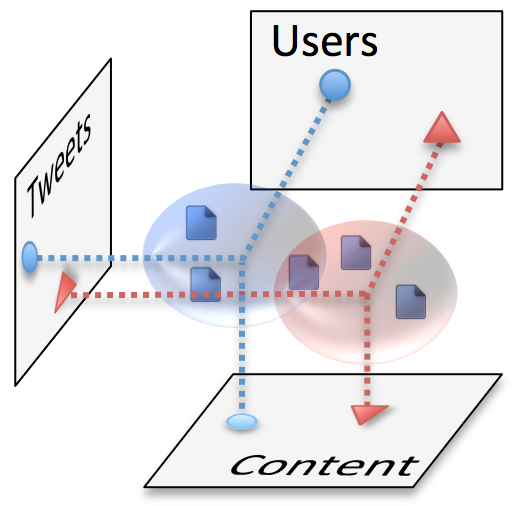}
}
\caption{Left: Relations among protomemes, tweets, users, and tweet content. Right: Projections of protomemes onto spaces based on their tweet, user, and content features that inform corresponding similarity measures.}
\label{fig:protomeme_space}
\end{figure*}

Hereafter we will refer to instances of these features as \emph{entities}.
Consider the tweet \textsl{``@All\_4Given Gingrich: Romney Most Likely Nominee http://t.co/CectDLni \#All4Given''} as an example. This tweet contains four entities: the hashtag \texttt{\#All4Given}, the mention \texttt{@All\_4Given}, the URL \url{http://t.co/CectDLni}, and the phrase \texttt{gingrich romnei most like nomine}. In the following we will denote as \emph{protomeme} the set of all tweets that contain a specific entity. Think of a protomeme as a primitive meme. One consequence of this representation is that protomemes overlap; the tweet in the above example belongs to four protomemes.

Additionally, any tweet is accompanied by a plethora of \emph{metadata}, such as author information (e.g., username, number of tweets and followers, self-reported user location), temporal and geographical information (e.g., timestamp and latitude/longitude coordinates of the tweet), retweet information, and so on. The set of features could be expanded to exploit such metadata. 

Protomemes leverage the \emph{wisdom of the crowd}~\cite{golder2006usage,mika2007ontologies}: users exploit content features that allow for the effective identification of discussion topics (hashtags), ongoing conversations (mentions), or external resources (URLs). Thus, by adopting protomemes we intend to alleviate the problem of text sparsity, which has proven to hinder the application of topic modeling techniques to Twitter~\cite{hong2010empirical}. 

Moreover, protomemes can aid in the task of clustering messages in a streaming scenario, as they are easily extracted in real time by defining a set of matching rules, such as regular expressions. Incoming tweets are seamlessly added to existing protomemes, which form natural initial tweet clusters. Protomemes therefore provide an efficient pre-clustering strategy to aggregate messages. 

Our social media clustering framework uses protomemes as the fundamental units. Natural similarity measures can be defined over the protomemes (sets of tweets), to aggregate related protomemes into broader memes. 

\subsection{Similarity measures} 
\label{sub:measures}

Fig.~\ref{fig:protomeme_space} (left) illustrates the mutual relations between protomemes, the tweets they contain, their content, the users who post them, and the underlying follower network. We can define similarity measures between protomemes by considering the projections of the protomemes onto spaces induced by these features, also depicted in Fig.~\ref{fig:protomeme_space} (right). 

Let us provide a few preliminary definitions. Let $P_{\ell}$ be a set of tweets, $U_{\ell}$ be the set of users that produced tweets in $P_{\ell}$ ($|U_{\ell}| \leq |P_{\ell}|$ because a user may post more than one tweet), and $W_{\ell}$ be the set of terms  obtained by concatenating all tweets in $P_\ell$. We can now  define a set of measures, which we will apply to compute the similarity between protomemes.

\begin{definition}[Common user similarity] $S_{u}$ between protomemes $P_i$ and $P_j$ is  the cosine similarity between their user frequency vectors
\begin{equation}
S_u(P_i, P_j) = \frac{\sum_{u \in U_i \cap U_j} P_{iu} P_{ju}}{\sqrt{\sum_{u \in U_i}P_{iu}^2}  \sqrt{\sum_{u \in U_j} P_{ju}^2}}
\label{eq:cus}
\end{equation}
where $P_{\ell u}$  is the number of times user $u \in U_{\ell}$ adopts protomeme $P_{\ell}$.
\end{definition}

\begin{definition}[Common tweet similarity] $S_t$ between protomemes $P_i$ and $P_j$ is  the cosine similarity between their (binary) tweet vectors
\begin{equation}
S_t(P_i, P_j) = \frac{|P_{i} \cap P_{j}|}{\sqrt{|P_{i}|} \sqrt{|P_{j}|}}.
\label{eq:cts}
\end{equation}
\end{definition}

\begin{definition}[Content similarity] $S_c$ between protomemes $P_i$ and $P_j$ is the cosine similarity between their TF-IDF vectors
\begin{equation}
S_c(P_i, P_j) = \frac{\sum_{w \in W_i \cap W_j} P_{iw} P_{jw}}{\sqrt{\sum_{w \in W_i}P_{iw}^2}  \sqrt{\sum_{w \in W_j} P_{jw}^2}}
\label{eq:cds}
\end{equation}
where $P_{\ell w}$ is the TF-IDF weight assigned to term $w \in W_{\ell}$.
\end{definition}

%We wish to introduce a fourth similarity measure rooted on the assumption of an underlying social network of users whose interaction allows for the propagation and the involved in a set of tweets (see illustration of follower network in Fig.~\ref{fig:protomeme_space}). 
Since we cannot assume that information about the follower social network is available to the clustering algorithm, let us exploit mention and retweet metadata as a proxy for the underlying network structure to define a forth similarity measure. Let $N_{\ell} = U_{\ell} \cup M_{\ell} \cup R_{\ell}$ be the diffusion set of $P_{\ell}$, where $M_{\ell}$ is the set of users mentioned in tweets in $P_{\ell}$, and $R_{\ell}$ is the set of users who have retweeted posts in $P_{\ell}$.\footnote{Note that $R_{\ell}$ is not necessarily a subset of $U_{\ell}$ when only a sample of the tweets are considered in the stream; the sample may include a retweeted message but not the original one.} 
%The diffusion set is a subset of the neighbors of users involved in the protomeme on the mention and retweet networks. 

\begin{definition}[Diffusion similarity] $S_d$ between protomemes $P_i$ and $P_j$ is the cosine similarity between their diffusion (binary) vectors
\begin{equation}
S_d(P_i, P_j) = \frac{|N_{i} \cap N_{j}|}{\sqrt{|N_{i}|} \sqrt{|N_{j}|}}.
\label{eq:cns}
\end{equation}
\end{definition}

\subsection{Combinations}
\label{sub:combinations}

There are many ways to combine different similarity measures. One of the goals of our experimentation will be to explore how different similarity measures and combinations of them affect the quality of the meme clusters. In the following we introduce two different methods incorporated in our framework.

The \emph{pairwise maximization strategy} aims at choosing the measure that provides the highest value every time we compute the similarity between two protomemes. The rationale is to capture the characteristics of a particular pair of protomemes; the relatedness of two protomemes may be best described by their content while that of two other protomemes may be more obvious by considering users, say. 
Given a set of similarity measures $S_1, \dots, S_n$, the \emph{maximum pairwise similarity} is formally defined as 
\begin{equation}
MAX(P_i, P_j) = \max_k \{S_k(P_i, P_j)\}.
\label{eq:maximization}
\end{equation}

A second approach is a \emph{linear combination} of similarity measures, extending the  idea of averaging~\cite{Qazvinian12,reisinger2010multi}. 
Formally, 
\begin{equation}
\mathcal{L}(P_i, P_j) = \sum_k \omega_k S_k(P_i, P_j)
\label{eq:combination}
\end{equation}
with the constraint that $\sum_k \omega_k = 1$,  allowing for a normalized combination such that $\mathcal{L}(P_i, P_j) \in [0,1]$ (assuming $ \forall k \; S_k \in [0,1]$).
The set of parameters $\omega_1, \dots, \omega_n$ introduces an $(n-1)$-dimensional parameter space whose exploration is instrumental for understanding what combinations of similarity measures provide the best performance in terms of clustering. This aspect will be investigated in detail in our experiments. %\S~\ref{sub:parameter_space}.

\subsection{Clustering algorithms} 
\label{sub:clustering}

The ideal clustering framework should allow for the detection of memes (clusters) at different levels of granularity --- in some cases one might prefer small clusters of tightly related protomemes, in other cases a smaller number of broader groups may be required. \emph{Hierarchical clustering algorithms} are naturally designed to span a range of granularities. On the other hand, if the desired number of clusters is known in advance, \emph{K-means} offers an efficient alternative. Since different clustering algorithms work best for different tasks, we will evaluate whether hierarchical or K-means clustering is better suited for our task. % (\S~\ref{sub:results}). 
Of course there are many other clustering methods; we only consider these two widely adopted techniques as we are more interested in the role of the protomemes and various similarity measures in determining cluster quality.\footnote{In the next section we extensively discuss criteria for determination of cluster quality.} Once a set of objects to cluster (tweets or protomemes) and similarity measures among objects are defined, it is possible to apply any off-the-shelf clustering algorithm --- or to design new ones. 

Both algorithms can appropriately work in our platform, taking as input the protomemes produced in the pre-clustering detection system from the datastream. In our system the clustering is intended as an \emph{asynchronous}, \emph{off-line} process, that at predetermined time intervals analyzes and clusters the amount of protomemes captured in a recent time window (say, the last hour of data). New tweets incoming from the social stream can be assigned to the existing clusters via the protomeme pre-clustering processing until the next execution of the clustering algorithm will further refine the existing clusters, exploiting additional data points and disregarding  protomemes not observed in recent data. 

Regarding hierarchical clustering, we adopt the \emph{average-linkage} method to determine the similarity among clusters. The similarity between two clusters is simply the average pairwise similarity among all protomemes belonging to them. This approach alleviates the sensitivity of the algorithm in the presence of outliers. The similarity matrix among all protomemes is computed once at the start.

\begin{figure}
	\includegraphics[width=\columnwidth]{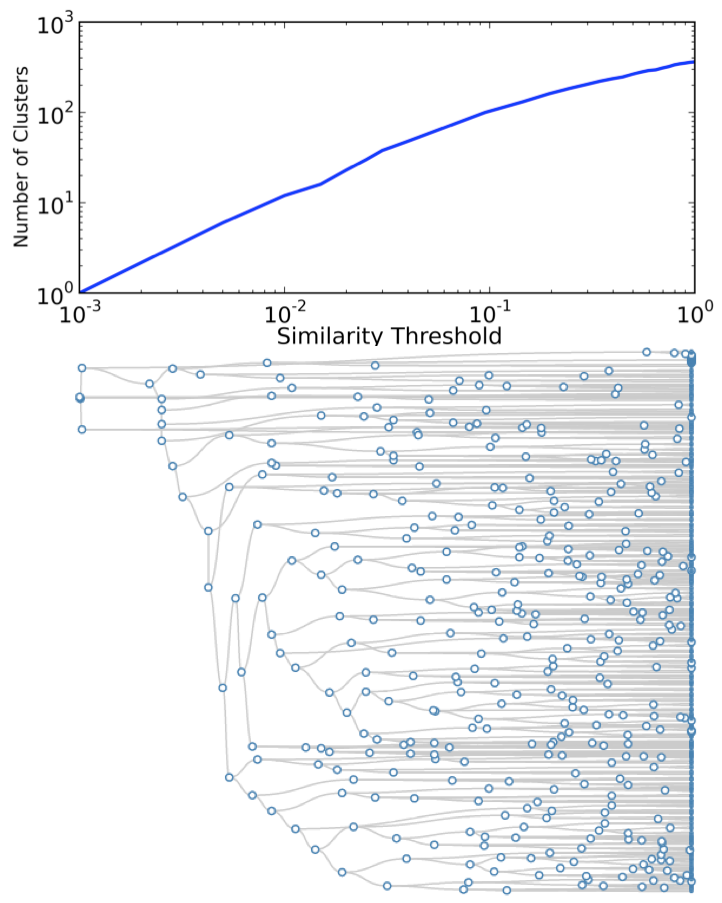}	
	\caption{Number of clusters (top) as a function of the similarity threshold used to cut the dendrogram (bottom) in hierarchical clustering.}
	\label{fig:dendrogram_sim}
\end{figure}

Once the dendrogram has been generated by the hierarchical clustering algorithm, a dendrogram cut is applied by picking a similarity threshold $\tau$. This process  allows for tuning the performance of hierarchical clustering to find the correct trade-off between number and size of clusters, as shown in Fig.~\ref{fig:dendrogram_sim}. In our experimental trials we will show how the number of clusters (determined by the choice of $\tau$) affects the quality of the clustering solution. While K-means clustering requires to select the number of clusters in advance, one can run K-means for different numbers of clusters, exploring a similar range of solutions. 

%####################
\section{Evaluation} 
\label{sec:experiments}

In this section we discuss a systematic evaluation process carried out to assess performance in the meme clustering task. First we describe a dataset adopted as ground truth in our evaluation. Then we introduce a quality metric, motivating its adoption and giving some intuition for how it works. 
%Finally, we discuss the experiments to evaluate the protomeme pre-clustering process and the various similarity measures and their combinations. 

\subsection{Ground truth dataset}

For evaluation purposes, we used an existing dataset of hand-curated tweets as ground truth. %\footnote{Unfortunately, the Twitter terms of service prevent us from sharing the dataset.} 
Tweets about political news regarding the US presidential primaries were collected during April 2012 using the Twitter APIs (\url{dev.twitter.com/docs/api}). To verify the completeness of the dataset, we compared it with Twitter's \emph{gardenhose} sample, which collects about 10\% of all public tweets. We observed that the number of tweets in the dataset was roughly ten times larger than those in the gardenhose data. This suggests that the dataset includes nearly all of the tweets that were published about these topics during the observation period. The dataset comprises of 5,523 tweets containing 2,866 URLs, 2,780 hashtags, and 1,848 mentions. 
Note that exact duplicates and retweets were removed from the dataset. This possibly penalizes the performance of the clustering algorithms by biasing the similarity measures. For example, retweet information is not available when computing diffusion similarity. 

%\subsection{Cluster Annotations}

We annotated the set of tweets in two steps. 
We first manually reviewed all tweets in the dataset identifying topics consisting of at least three tweets. 
Then we labeled each tweet with one or more topics. 
Our annotations resulted in a set of clusters corresponding to 26 topics.
Some examples of tweets contained in the dataset and relative cluster assignments are reported in Table \ref{tab:examples}.
%In light of this fact, during our experimental evaluation we will deal with this class in two different ways, first we include it, and then we filter it out during a data pre-processing step in order to assess how it affects the performance of our system.

Given the brevity of tweets, most (92.1\%) discuss only a single topic.
However, 7.9\% of the tweets were assigned to more than one cluster.
% as shown in Table \ref{tbl:clustercount}. 
We will exploit this information during our evaluation to assess whether our clustering framework is able to capture such overlap. 
Table \ref{tbl:classes} reports the composition of the clusters obtained after the manual annotation. 
%For each of them we also report the fraction of tweets non-uniquely assigned to that cluster (i.e., the \emph{overlap ratio}).

%\begin{table}[!h] \centering \small
%\caption{Number and percentage of tweets that belong to only one, two, or more clusters.}
%\begin{tabular}{|c|c|c|}
%\hline
%\multicolumn{3}{|c|}{Number / \% of tweets}\\ \hline
%One cluster & Two clusters &  More clusters \\ \hline
%11,724	&	437 &	1	\\
%96.4\%	&	3.6\% &	0.008\%	\\
%\hline
%\end{tabular}
%\label{tbl:clustercount}
%\end{table}

\begin{table*}[!t] \centering \footnotesize
\caption{Examples of thematically related tweets manually assigned to one of the 26 clusters (``Santorum out the race'').}
\begin{tabular}{l} 
\hline
Rick Santorum ends presidential campaign after conceding to Mitt Romney in phone call - The Ticket - Yahoo! News http://t.co/L6IYHt6d \\
Santorum suspends campaign, clearing Romney's path http://t.co/HC3XtptZ \#cnn \\
\#BREAKING: Rick \#Santorum suspends his campaign for president, making Mitt \#Romney likely Republican nominee.\\
MITT ROMNEY EXPRESS: Is Rick Santorum In Or Out Of The 2012 Race? http://t.co/zva9ZK3i	\\
Rick Santorum quits campaign to leave field clear for Romney http://t.co/oKVBUYo3 \#News \#CNN \#Politico	\\
\hline
\end{tabular}
\label{tab:examples}
\end{table*}

\begin{table}[!t] \centering \footnotesize
\caption{Composition of the clusters obtained by manual labeling. The overlap ratio is the percentages of tweets in a cluster that also belong to at least another cluster.}
\begin{tabular}{crr|crr}
%\hline
%\multicolumn{6}{|c|}{Label / no. tweets / overlap ratio}\\
\hline
Cluster & Tweets & Overlap  & Cluster & Tweets & Overlap  \\
 &  &  \multicolumn{1}{c|}{ratio} &  &  &  \multicolumn{1}{c}{ratio} \\
\hline 
1	&	1,654	&	16.14\%	&	14	&	57	&	28.07\%	\\
2	&	1,522	&	13.14\%	&	15	&	54	&	35.18\%	\\
3	&	405	&	29.62\%	&	16	&	49	&	24.49\%	\\
4	&	376	&	6.64\%	&	17	&	43	&	0.00\%	\\
5	&	355	&	14.92\%	&	18	&	41	&	4.88\%	\\
6	&	343	&	17.20\%	&	19	&	35	&	20.00\%	\\
7	&	245	&	8.57\%	&	20	&	29	&	0.00\%	\\
8	&	230	&	9.56\%	&	21	&	27	&	22.22\%	\\
9	&	128	&	3.91\%	&	22	&	21	&	14.28\%	\\
10	&	97	&	18.55\%	&	23	&	20	&	0.00\%	\\
11	&	86	&	15.11\%	&	24	&	4	&	25.00\%	\\
12	&	71	&	0.00\%	&	25	&	4	&	0.00\%	\\
13	&	63	&	6.35\%	&	26	&	3	&	0.00\%	\\
\hline
\end{tabular}
\label{tbl:classes}
\end{table}

\subsection{Evaluation metric}

To assess the quality of the clusters, we adopt a measure based on \emph{Normalized Mutual Information} (\emph{NMI}) \cite{danon2005comparing}. The NMI assumes the availability of a {\em ground truth} that represents the correct clusters. Let $A$ be the correct cluster assignment, and suppose that it contains $c_A$ clusters. Let $B$ be the output of a clustering algorithm operating on the same data and producing $c_B$ clusters. We can define a $c_A \times c_B$ \emph{confusion matrix} $\mathbf{N}$, whose rows correspond to the clusters in $A$ and whose columns represent clusters in $B$. Each entry $N_{ij}$ of this confusion matrix reports the number of elements of the correct $i$-th cluster that happen to be assigned to the $j$-th cluster by the clustering algorithm. The \emph{Normalized Mutual Information} is defined as

{\small
\[
%\begin{equation} 
NMI(A,B) = \frac{-2 \displaystyle\sum_{i = 1}^{c_A}\sum_{j = 1}^{c_B} N_{ij} \log \left(\frac{N_{ij}N}{N_{i\cdot}N_{\cdot j}} \right)}{\displaystyle\sum_{i=1}^{c_A}N_{i \cdot} \log \left(\frac{N_{i \cdot}}{N}\right) + \displaystyle\sum_{j=1}^{c_B}N_{\cdot j} \log \left(\frac{N_{\cdot j}}{N}\right)}
\label{eq:nmi}
\]
%\end{equation}
}
where $N_{i \cdot}$ (resp., $N_{\cdot j}$) is the sum of the elements in the $i$-th row (resp., $j$-th column) of the confusion matrix, and $N$ is the sum of all elements of $\mathbf{N}$. The output of this measure is normalized between zero (when the clusters in the two solutions are totally independent), and one (when they exactly coincide). Therefore, the higher the value of NMI, the better the quality of the clusters found by the algorithm.

Measures based on mutual information have been shown to best capture different facets of a clustering process, such as how well a clustering algorithm reflects the number, size, and composition of clusters with respect to the ground truth, as opposed to some other widely used measures, which may produce biased evaluations \cite{meilua2007comparing}. Our investigation with accuracy, precision, recall, and $F_1$ confirmed the limitations of these measures, all of which report indistinguishable results due to the dominance of true negatives. Purity, on the other hand, is by definition biased toward rewarding the presence of tiny clusters, which tend to be pure. For these reasons, NMI has been recently adopted in the evaluation of tasks such as event detection in social media~\cite{becker2010learning}. 

The previous definition does not work well in the case of overlapping clusters. We therefore adopt in our evaluation a variant called LFK-NMI, which accounts for overlapping clusters. Details on the formulation of LFK-NMI are outside the scope of this paper, and can be found in the paper by Lancichinetti \emph{et al.}~\cite{lancichinetti2009detecting}. 

\section{Results} 
\label{sub:results}

There are many potential configurations of our clustering framework; next we present experiments designed to evaluate several aspects of meme clustering.

\subsection{Hierarchical vs. K-means clustering}
\label{sub:algorithms}

\begin{figure}[!t] \centering
	\includegraphics[width=.9\columnwidth]{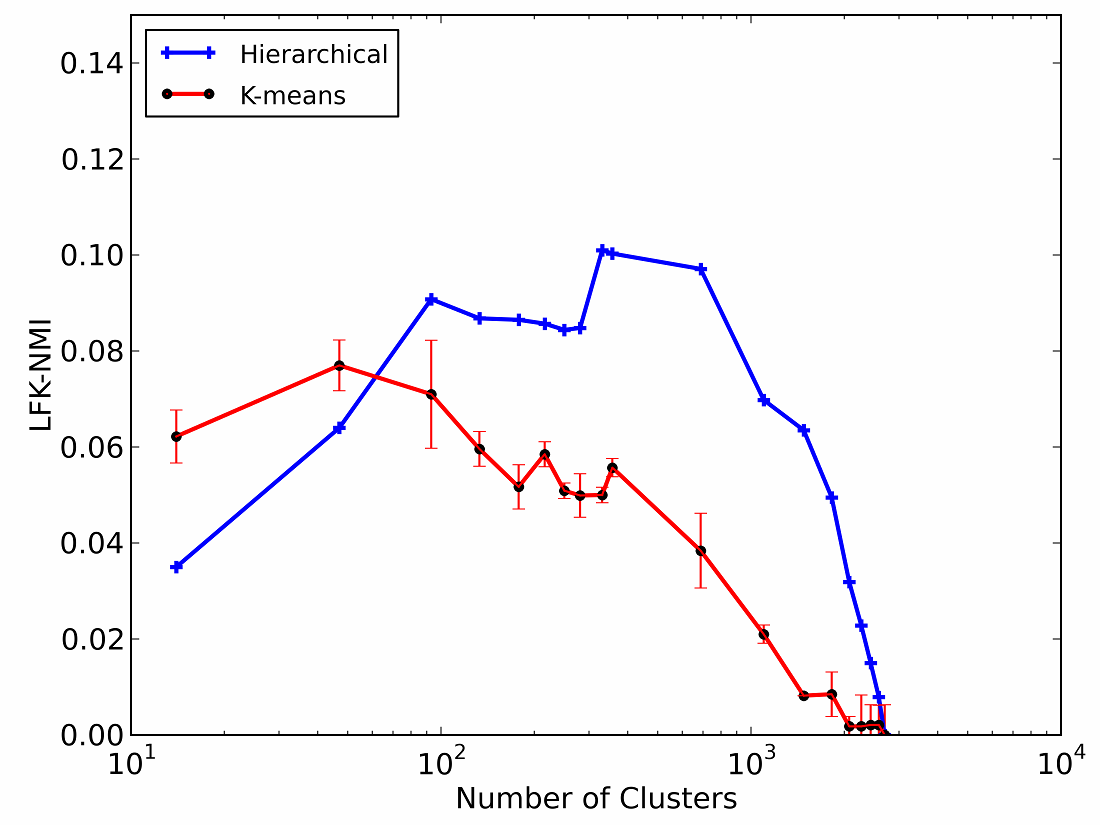}	
	\caption{Performance comparison between hierarchical and K-means clustering. The task for this experiment is that of clustering protomemes using a single similarity measure, namely content similarity. The error bars for K-means are standard errors based on five runs for each number of clusters.}
	\label{fig:kmeans_vs_hierarchical}
\end{figure}

The first experiment  aims at choosing one clustering algorithm. Fig.~\ref{fig:kmeans_vs_hierarchical} compares hierarchical and K-means clustering algorithms.   We report the value of LFK-NMI as a function of the number of clusters, obtained with each of the clustering methods. By varying the similarity threshold $\tau$ in hierarchical clustering, we obtained partitions with different numbers of clusters (cf.~Fig.~\ref{fig:dendrogram_sim}). We then ran K-means for each of the corresponding numbers of clusters.  

K-means generally performs  well at finding a very small number of clusters, but the quality of discovered clusters quickly deteriorates when the number of clusters increases. Hierarchical clustering outperforms K-means over a broad range of values for the number of clusters, and also achieves a better overall cluster quality. 
%This suggests an advantage of hierarchical clustering when the aim is that of finding small clusters. This might be useful in particular when looking at the early inception of memes.

While the experiment reported in Fig.~\ref{fig:kmeans_vs_hierarchical} is based on protomemes and content similarity, we systematically explored other measures. All experiments provided the same verdict: hierarchical clustering outperforms K-means in our meme clustering task on Twitter. As a result, we employ hierarchical clustering in the rest of our evaluation.

\subsection{Experimental setup}
\label{sub:exp-protomemes}

Our second and more important experiment aims at assessing whether protomemes convey a concrete advantage in the task of clustering memes in social media. To this purpose we compare two different configurations of our clustering framework that operate with protomemes against two configurations that operate directly on individual tweets. 
%We explored various techniques in the literature, starting from simple string edit distance measures, to those based on as \emph{n}-grams and vector space models. However, neither simple nor complex text-based measures achieved satisfactory performance 
%
%The scope of this experiment is twofold. First, as mentioned above, we address the question of whether protomemes are beneficial to clustering. Then, we explore how different configurations of our clustering framework perform against each other, in order to assess what similarity measures and what features of data best capture the nuances of the rumors we are dealing with in our current evaluation.
%
%Without protomemes, one can compute content similarity directly between the textual features of individual tweets. 
%We explored various techniques in the literature, starting from simple string edit distance measures, to those based on as \emph{n}-grams and vector space models. However, neither simple nor complex text-based measures achieved satisfactory performance 
%
The description of these four configurations follows.

\textsl{Baseline:}
	This configuration is a straightforward implementation of hierarchical clustering of simple tweets.
	The similarity measure adopted to compare tweets and aggregate them is content similarity based on TF-IDF.
%, which in our testing has proven to work better than other vector-based techniques (such as TF-IDF) or $n$-grams.

\textsl{Baseline+Followers:}
	This configuration is inspired by the event detection system recently proposed by Aggarwal and Subbian~\cite{aggarwal2012event}, namely a tweet clustering algorithm based on both content- and network-based features.
	The main limitation of this approach is that it relies on the full knowledge of the follower network of all users present in the dataset. 
	Such  information is very time-consuming to obtain and this task seems unfeasible in real-time, streaming scenarios.
	To compute tweet similarity, the authors adopt TF-IDF, and we reproduced this choice in our implementation. Follower similarity is implemented using Eq.~\ref{eq:cns} but with actual follower sets; this is a variation of the original formulation, which is not applicable in our framework. The two similarity measures are linearly combined with equal weights, and used with hierarchical clustering (note that hierarchical clustering provides better performance than K-means in this case as well.)
	
\textsl{MAX:}
	In this configuration we adopt protomemes as the atoms of our clustering algorithm. 
	We use all four similarity scores defined above ($S_t$, $S_u$, $S_c$, and $S_d$), and combine them via the maximum pairwise similarity (Eq.~\ref{eq:maximization}).

\textsl{Linear combination:}
	The last configuration also exploits all four similarity measures, combined by way of a linear combination (Eq.~\ref{eq:combination}). We discuss how to determine the parameters of the linear combination later in this section.

\begin{figure}[!t] \centering
	\includegraphics[width=.9\columnwidth]{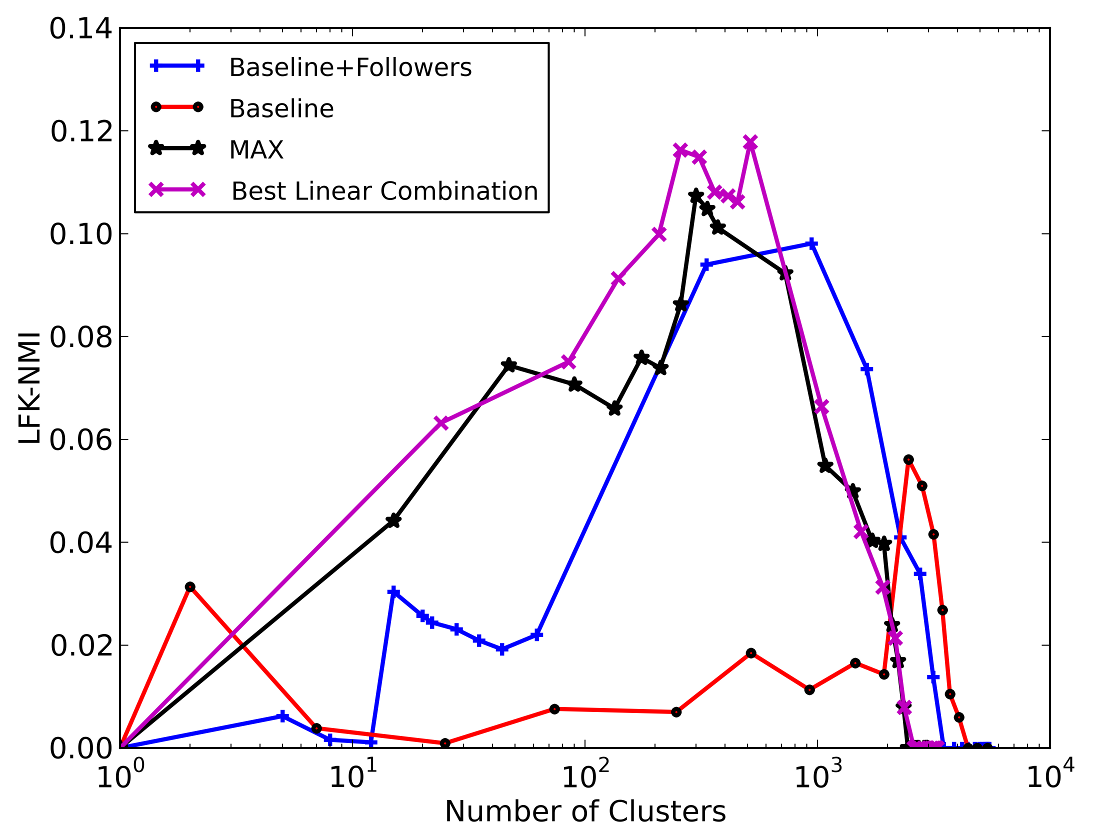}
	\caption{Performance of different clustering algorithms, as a function of the number of clusters.}
	\label{fig:lfk_nmi_results_noBundle}
\end{figure}

Fig.~\ref{fig:lfk_nmi_results_noBundle} presents a comparison between the performance of the four configurations. 
The baseline achieves its best performance for a number of clusters that is comparable to the number of tweets, which is not very helpful in our setting. This is understandable since either the tweets are very similar, or due to the sparsity of the text, their similarity is likely close to zero. The Baseline+Followers algorithm performs better, achieving higher quality and with fewer clusters. This configuration, however, might not be viable for our task in the streaming scenario. 
The MAX strategy obtains its best performance for a much smaller number of clusters, even outperforming Baseline+Followers when the number of clusters is closer to that in the ground truth. This is remarkable given that MAX does not have access to the full follower network. We interpret these results as evidence that protomemes provide a significant advantage.
This advantage becomes further evident considering the performance of the linear combination. This configuration provides a slight improvement over the MAX strategy, although the parameters are optimized with knowledge of the ground truth. 
%The weights for the linear combination are obtained with an optimization procedure detailed next.

%\subsection{Exploring the parameter space} 
%\label{sub:parameter_space}

%%%%% REDO FIGURES WITH SHOWING SAME OPTIMAL WEIGHTS AS MENTIONED IN TEXT
\begin{figure}[!t] \centering
\includegraphics[width=.9\columnwidth]{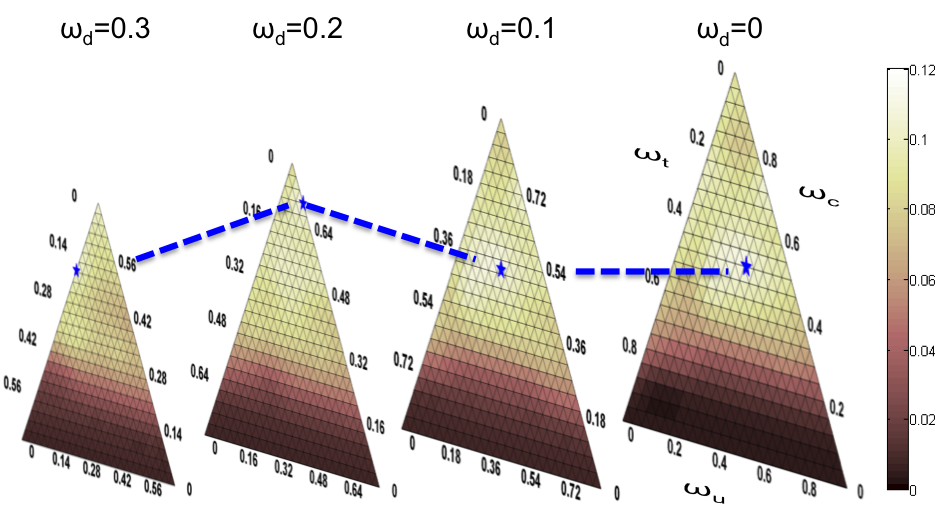}
\caption{Slices of the 3-simplex showing LFK-NMI values originated by the linear combination of the four similarity measures: \emph{common user similarity} ($S_u$) on the bottom edge, \emph{content similarity} ($S_c$) on the right edge, \emph{common tweet similarity} ($S_t$) on the left edge, and \emph{diffusion similarity} ($S_d$) across the slices. The combination yielding highest LFK-NMI in each slice is highlighted. 
%We also display the trajectory of the optimal solution across slices to appreciate the stability of results. 
} 
\label{fig:parameter_space}
\end{figure}

%The following experiment is devoted to exploring the $4$-dimensional parameter space introduced by adopting a linear combination of all four similarity measures, as seen in Equation \ref{eq:combination}. While computing the similarity among protomeme clusters, the linear combination accounts for different measures at the same time, exploiting different weights according to the parameters. We explore the parameter space to assess what combination allows for the best clustering performance.

To determine the weights of the linear combination of the four similarity measures (Eq.~\ref{eq:combination}), we used a greedy optimization procedure. 
%there exist two possible ways to proceed. One direction is to fix the clustering method, and 
We ran the hierarchical clustering algorithm for each set of weights in a 3-dimensional simplex with step 0.1, resulting in 286 parameter configurations. 
% times by varying the parameters of the linear combination of $S_u$, $S_c$, $S_t$ and $S_d$, and the similarity threshold. 
%Then, compute the quality of each clustering solution according to LFK-NMI with respect to the ground truth. 
For each parameter set we found the best LFK-NMI value, irrespective of the number of clusters. We finally selected the best overall setting.
%, and the corresponding threshold. 
%This procedure is effective but computationally expensive. Another efficient weay to deal with the problem is as follows: one can pick the threshold that produces the optimal value based on the pairwise maximization strategy and explore the parameter space only in the neighborhood of that maximum. Once identified the best linear combination, one can use the obtained values to run the algorithm on the entire threshold range, finally obtaining the best solution. In our experimentation, the latter procedure allows for a rough 10\% performance improvement simply discovering a local optimum starting from the global optimum discovered by the maximization strategy. Interestingly enough, in our analysis that compares these two strategies, we observed comparable performance, with the latter approach being an order of magnitude less expensive than the former one.

Fig.~\ref{fig:parameter_space} shows that optimal solutions (high LFK-NMI values) are provided by non-trivial combinations of the four parameters.
%In detail, we show a 3-simplex sliced across the plane created by three parameters (i.e., $S_t$, $S_c$, and $S_u$) and orthogonal along the forth parameter ($S_d$). 
The highest LFK-NMI peak is obtained with the following weights: $\omega_t=0.0$, $\omega_c=0.7$, $\omega_u=0.1$, $\omega_d=0.2$. Common tweet similarity does not contribute to this particular configuration, although it is to be noted that other parameter settings yield higher LFK-NMI for different values of the number of clusters (for example, another configuration with $\omega_t=0.2$ yields compareble LFK-NMI for fewer clusters.)
%\footnote{Note that in this specific case the algorithm benefits little from the diffusion similarity because no retweet information is available in the dataset used for evaluation. We expect such measure to contribute much better in other scenarios.}
%This result suggests that in the future we can further improve performance of our clustering framework by improving the learning strategy and possibly by introducing additional similarity measures.

\subsection{Cross-validation} 
\label{sec:cross-validatin}

The optimization procedure outlined above for tuning the parameters of the linear combination of similarity measures uses knowledge about the ground truth labels for both learning the weights and evaluating the quality of the clusters. This runs the risk of over-fitting on a training set. To assess the robustness of our performance evaluation, we performed a cross-validation analysis.

%In the last experiment we investigate the consistency and robustness of the presented framework by means of cross-validation analysis. 
%The procedure that allows us to achieve the better performance by means of a linear combination of similarity measures, involves a step of parameter learning. In fact, in the previous experiment we used a manually-curated dataset as ground truth to estimate the quality of clusters
We opted for 5-fold cross-validation so that the test sets would not be too small for cluster quality evaluation. 
%rather than the standard 10-fold one due to the problem of class imbalance: according to Table~\ref{tbl:classes}, two orders of magnitude separate the sizes of the smallest and largest topics in the ground truth. 10-fold cross-validation could potentially allow for over-sampling the largest classes in the folds generated first, leaving the least represented classes being over-represented in the latest folds. This issue is mitigated by reducing the number of folds to 5. Another shrewdness adopted during the creation of the 5 folds is to 
We randomly assigned each tweet to one of the 5 folds, 
%rather than purely sampling at random five times 20\% of the datapoints. This 
guaranteeing equal fold sizes and unbiased distribution of topics across each fold, in spite of the considerable class imbalance (cf.~Table~\ref{tbl:classes}). For each iteration, we used the combination of 4 folds as training data (to optimize the weights as described in the previous subsection), and the remainder 20\% of the tweets as test set (to measure clustering quality). Performance of the clustering algorithm was computed after splitting the ground truth topics consistently with the generated folds. Results were finally averaged across the 5 runs of the cross-validation test.  

\begin{figure}[!t] \centering
	\includegraphics[width=.9\columnwidth]{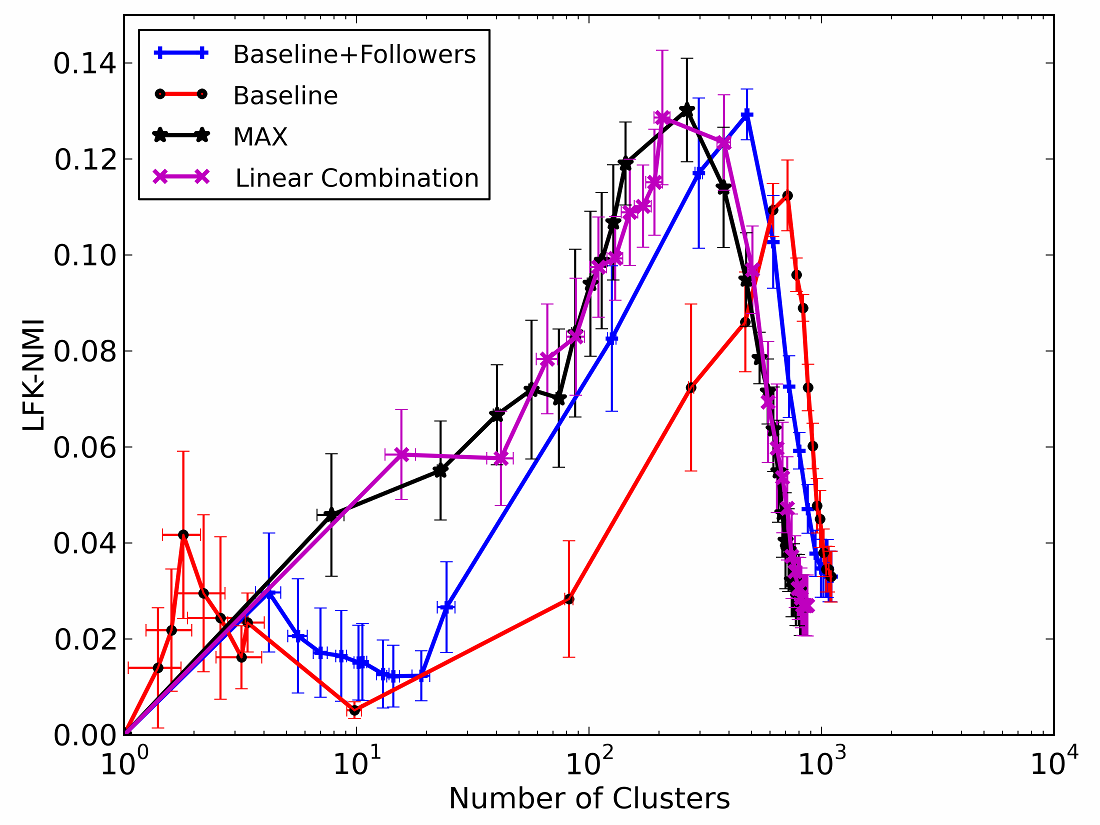}
	\caption{Results of 5-fold cross-validation on our dataset. Error bars represent standard errors on the number of clusters and LFK-NMI across folds, for each dendrogram cut.} 
\label{fig:cross-validation}
\end{figure}

Fig.~\ref{fig:cross-validation} shows the results of our cross-validation analysis. 
%The linear combination reports standard error bars both on the x and y axis. The ones on the x axis explain, given a threshold used to cut the dendrogram generated by the hierarchical clustering, what is the variation on the number of clusters obtained with using of the five folds. The bars on the y axis report the variation on the achieved LFK-NMI for the given solution according to the 5 iterations of cross-validation. 
It is worth noting that the test sets contain only 20\% of the data, and therefore the number of clusters is smaller than that reported in Fig.~\ref{fig:lfk_nmi_results_noBundle}, where performance was assessed on the entire dataset. To obtain a meaningful comparison, we evaluated the other algorithms on the same test sets. They perform relatively better on this easier clustering task, as reflected in higher values of LFK-NMI compared to Fig.~\ref{fig:lfk_nmi_results_noBundle}.

The cross-validation analysis demonstrates that the performance obtained with the learned linear combination of similarity measures is not statistically better than that obtained with the MAX strategy.
%From Fig.~\ref{fig:cross-validation} it is possible to observe that the linear combination strategy is very stable and produces the better performance both with respect to the quality of retrieved clusters and their number (the closest to the ground truth, the better). Also, note that here cross-validation servers two different purposes in the case of linear combination and in the others. In fact, in the former case cross-validation allows to investigate both whether the parameter learning process amounts for over-fitting on training data and also if the algorithm performance is consistent in respect to different chunks of data. In the other cases (i.e., pairwise maximization, baseline and baseline+followers) instead, there is no parameters learning involved, then the cross-validation assess only the resiliency of the algorithm with respect to different testing datasets.
Interestingly, we can achieve close to optimal performance without having to assume prior knowledge of the ground truth, which of course makes our clustering algorithm more amenable to a realistic streaming scenario.
%In summary, these results indicate that the linear combination strategy shows better performance if compared against other methods and a great resilience with respect to data. For such reasons, in the future we plan to devote further attention to design additional measures to incorporate in our model and better parameter learning methodologies.

\section{Related work} 
\label{sec:related-work}

This work, to the best of our knowledge, is the first to formalize the problem of clustering memes in online social media. Recent literature has discussed related problems, such as the identification of topics or memes  \cite{leskovec2009meme,xie2011visual,simmons2011memes} or emerging events in social streams \cite{sayyadi2009event,cataldi2010emerging,mathioudakis2010twittermonitor,marcus2011twitinfo,lehmann2012dynamical}.

Leskovec \textit{et al.}~\cite{leskovec2009meme} presented \textsl{memetracker}, a platform that tracks memes produced in online media such as mainstream news sites and Weblogs. \textsl{Memetracker} can group together short, distinctive phrases that act as signatures of specific topics and identify small variations of them. This creates groups of news on related topics that can be tracked over time to define patterns of diffusion in the news cycle. \textsl{Memetracker} identifies and aggregates disjoint memes on the basis of textual similarity but no systematic evaluation of the quality of the retrieved memes is provided. Our work instead focuses on the assessment of the quality of the meme clustering process, and allows for overlapping memes.

The problem of tracking news for meme extraction has been tackled also by Simmons \textit{et al.}~\cite{simmons2011memes}. Based on the \textsl{memetracker} dataset, they investigated the extent to which information evolves and mutates due to collective processing of social media users. While defining protomemes, we rooted our work on the findings of both Leskovec \emph{et al.}~\cite{leskovec2009meme} and Simmons \emph{et al.}~\cite{simmons2011memes}, expanding on the aggregation of meme variants based not only on textual similarity, but also on other network and meta-data features.

Our framework shares some similarities also with another line of research on event detection systems. 
Aggarwal and Subbian~\cite{aggarwal2012event} presented a clustering algorithm that exploits both content- and network-based features to detect events in social streams. We adapted their algorithm to work in the context of meme clustering. Unfortunately, the algorithm assumes a preexisting knowledge about the follower network of Twitter users. In a streaming scenario, such information is expensive to get, especially when encountering popular users. In our framework, we proposed to rely on mention and retweet diffusion sets, which can be inferred in real-time from the observed data. 
Also, we achieved better performance by pre-clustering based on protomemes and relative similarity measures. 

Agarwal \emph{et al.}~\cite{agarwal2012real} recently proposed a graph-based algorithm for the real-time discovery of clusters in dynamic networks. The strategy is based on the discovery of dense clusters on the inferred graph of correlated keywords, extracted from tweets in a given time-frame. This method relies on the adoption of the \emph{short cycle property} that allows to find a local approximate solution. Performance of the system has been tested by using a simulated stream of tweets based on events reported by Google news in a given period, yielding high precision/recall in the task of identifying the largest events. 

Concluding, Becker \emph{et al.}~\cite{becker2011beyond} presented an event classification system designed for Twitter. The authors used temporal features in addition to social and topical ones. These features are adopted to train a classifier that consumes manually annotated clusters of data points representing specific events on Twitter. The best performance is provided by SVM, being a Naive Bayes classifier used as baseline. The results provided by the authors are promising and represent a starting point in the task of classifying different types of memes in Twitter.

%####################
\section{Conclusions} 
\label{sec:conclusions}

In this work we formalized the problem of clustering memes from social streams such as Twitter, and we presented a framework to deal with this task.
Our clustering framework adopts a novel pre-clustering procedure, namely protomeme detection, aimed at identifying atomic tokens of information inside tweets.
Due to its efficiency, this strategy should be particularly well suited to work in streaming scenarios. Additional work will be needed to empirically confirm this conjecture. 

Several similarity measures among protomemes have been defined, leveraging various features including content- and network-based ones, to build clusters of semantically and structurally related tweets.
The proposed diffusion similarity measure uses mention and retweet information, that can be reconstructed in real-time from the observed data, considering each protomeme diffusion set. 
We carried out a systematic evaluation showing the promising performance of the clustering framework, by using a manually-curated dataset as a ground truth. 
The best trade-off between quality, number, and size of clusters is obtained by pre-clustering using protomemes, and combining similarity measures exploiting heterogeneous features, with a simple pairwise maximization strategy. This approach performs as well as methods that assume prior knowledge of the data, and better than methods that assume knowledge of the underlying social network.
%non-trivial combinations of similarity measures, and a hierarchical clustering approach that allows for better determination of the composition of clusters. 

As for future work, we will extend the set of features incorporated by our clustering framework, considering for instance images. Furthermore, our preliminary analysis suggests that the introduction of time series as  features may yield significant performance improvements.

Our long-term plan is to integrate the meme clustering framework with a meme classifier to distinguish engineered types of social media communication from spontaneous ones. This platform will adopt supervised learning techniques to classify memes and determine their legitimacy, with the aim of early detection of attempts to spread misinformation and deceiving campaigns. The platform will be optimized to work with the real-time, high-volume streams of messages.

\section*{Acknowledgment}
The authors are grateful to Qiaozhu Mei, Sergey Malinchik, and Zhe Zhao for fruitful discussions.
This work is supported by NSF (grants CCF-1101743 and IIS-0968489), DARPA (grant W911NF-12-1-0037), and the McDonnell Foundation. The funders had no role in study design, data collection and analysis, decision to publish, or preparation of the manuscript.



%\bibliographystyle{IEEEtran}
{\footnotesize
\bibliographystyle{abbrv}
\bibliography{bibliography}  
}

\end{document}